\documentclass[12pt]{iopart}
\usepackage{graphicx}
\usepackage[numbers, sort&compress]{natbib}
\usepackage{iopams}
\begin{document}

\title[Multi-Task Learning for automated contouring and dose prediction in Radiotherapy]{Multi-Task Learning for Integrated Automated Contouring and Voxel-Based Dose Prediction in Radiotherapy}



\author{Sangwook Kim$^{1, 7}$, Aly Khalifa$^{1}$, Thomas G. Purdie$^{1, 2, 4, 8}$, and Chris McIntosh$^{1, 2, 3, 5, 6, 7}$}
\address{$^1$Department of Medical Biophysics, University of Toronto, Toronto, Canada}
\address{$^2$Princess Margaret Cancer Centre, University Health Network, Toronto, Canada}
\address{$^3$Toronto General Hospital Research Institute, University Health Network, Toronto, Canada}
\address{$^4$Princess Margaret Research Institute, University Health Network, Toronto, Canada}
\address{$^5$Peter Munk Cardiac Centre, University Health Network, Toronto, Canada}
\address{$^6$Department of Medical Imaging, University of Toronto, Toronto, Canada}
\address{$^7$Vector Institute, Toronto, Canada}
\address{$^8$Department of Radiation Oncology, University of Toronto, Toronto, Canada}
\ead{\{sangwook.kim,aly.khalifa,tom.purdie,chris.mcintosh\}@uhn.ca}


\vspace{10pt}
\begin{indented}
\item[]November 2024
\end{indented}

\begin{abstract}
Deep learning-based automated contouring and treatment planning has been proven to improve the efficiency and accuracy of radiotherapy. However, conventional radiotherapy treatment planning process has the automated contouring and treatment planning as separate tasks. Moreover in deep learning (DL), the contouring and dose prediction tasks for automated treatment planning are done independently.
In this study, we applied the multi-task learning (MTL) approach in order to seamlessly integrate automated contouring and voxel-based dose prediction tasks, as MTL can leverage common information between the two tasks and be able able to increase the efficiency of the automated tasks. We developed our MTL framework using the two datasets: in-house prostate cancer dataset and the publicly available head and neck cancer dataset, OpenKBP. Compared to the sequential DL contouring and treatment planning tasks, our proposed method using MTL improved the mean absolute difference of dose volume histogram metrics of prostate and head and neck sites by 19.82\% and 16.33\%, respectively.
Our MTL model for automated contouring and dose prediction tasks demonstrated enhanced dose prediction performance while maintaining or sometimes even improving the contouring accuracy. Compared to the baseline automated contouring model with the dice score coefficients of 0.818 for prostate and 0.674 for head and neck datasets, our MTL approach achieved average scores of 0.824 and 0.716 for these datasets, respectively.
Our study highlights the potential of the proposed automated contouring and planning using MTL to support the development of efficient and accurate automated treatment planning for radiotherapy.
\end{abstract}
\section{Introduction}

Radiotherapy is a critical component in cancer treatment, with the aim of precision and efficacy in targeting tumor tissues while sparing healthy ones. By automating certain components of radiotherapy, the advent of machine learning has promised to significantly advance the accuracy and efficiency of radiotherapy workflow~\cite{mcintoshnatmed}. Specifically, deep learning (DL) has been used to automate two key tasks: region of interest (ROI) contouring and dose prediction. Contouring ROIs in automated radiotherapy is crucial as it delineates the target tumor and surrounding organs at risk (OARs), enabling precise treatment planning and delivery. In terms of predicting dose distributions, utilizing DL can enhance the precision and efficiency of radiotherapy treatment planning by leveraging complex anatomical relationships to optimize clinical trade-offs in target coverage, dose conformity, and sparing healthy tissues.

However, because manual contouring is often required and is time-consuming, it represents a potential bottleneck in treatment planning when performed sequentially~\cite{mcintoshnatmed}. Thus, DL contouring is crucial in automated radiotherapy for efficiently and accurately delineating tumors and surrounding anatomy. Since the emergence of the U-Net architecture proposed by Ronneberger \etal, DL has enabled detailed and accurate delineation of anatomical structures in CT and MRI images~\cite{ronneberger2015u, isensee2021nnu_seg1, alzahrani2023geometric_seg2, ni2024generalizability_seg3}. The success of U-Net lies in its ability to capture context from a wide image area while maintaining high-resolution features, which is crucial for capturing the fine-grained features required in treatment planning. 

Owing to the ability of DL models in capturing a precise anatomical relationships, DL has also been widely used for predicting dose distributions in radiotherapy. Developing automated treatment planning includes training DL models using human contours and simulated dose information, that enables accommodating the unique complexities of individual patients. Although DL has enabled automated contouring and dose prediction in radiotherapy, studies thus far have treated these as separate tasks even though the two tasks are highly related. Particularly, studies have focused on developing separate dose prediction DL models that utilize human or DL contours as inputs~\cite{Nguyenfeasibilitysr, openkbp, openkbp_opt, gronberg2023deep_seq1, wen2023transformer_seq2, liu2021cascade_seq3, nguyenpmb3d_seq4, gao2023flexible_seq5, feng2023diffdp_seq6}. By utilizing sequential models, previous research lacks an investigation into the impact of simultaneously training for automated contouring and dose prediction using DL. Gu \etal conducted a notable study demonstrating how input contour information influences the development of separate DL dose prediction models~\cite{gu2023doseinputdata}. Specifically, they showed that generative adversarial networks can generate clinically acceptable dose distributions for head and neck cancer patients using only CT scans and ROIs. Importantly, their results indicated that DL models can be trained with minimal anatomical information on OARs.

Building on this, the work presented in this paper introduces our novel framework for DL contouring and dose prediction into an automated, integrated process. Our approach simultaneously optimizes model training for automated contouring and dose prediction, yielding enhanced performance compared to conventional sequential models trained on CT scans with separately extracted contour information. To achieve simultaneous training, we adopt multi-task learning, which is a branch of machine learning that exploits commonalities and differences across tasks~\cite{caruana1997multitask}. This approach excels in medical image analysis, where the interplay between anatomical structures, pathological features, and treatment planning elements can be exploited to enhance performance~\cite{haque2021generalized_mtl1, amyar2020multi_mtl2, chen2019multi_mtl3, moeskops2016deep_mtl4, zhao2023multi_mtl5, kim2024conversion_mtl6, maniscalco2024multimodal_mtl7, jiao2022mask}. Previously, Kim \etal introduced an attention-based multi-task learning model for automatic contouring and predicting dose distributions for prostate cancer and head and neck cancer radiotherapy~\cite{kim2023cross}. Another study by Jiao \etal introduced a multi-task learning generative adversarial network for automated dose prediction and contouring tumor mask. Both studies demonstrated that multi-task model can synergistically enhance the accuracy of both contouring and dose prediction by simultaneously identifying tumors while understanding surrounding critical structures.

To prove the efficacy of the multi-task learning model, we show in this work the performance improvement over sequential contouring and dose prediction models. We validate our approach using two different radiotherapy cancer treatment sites: prostate and head and neck cancer. To the best of our knowledge, this research is novel in terms of integrating two clinically imperative tasks in radiotherapy via multi-task learning. Although the study by Jiao \etal also demonstrated the feasibility of multi-task learning in radiotherapy treatment planning tasks for automated dose prediction and automated contouring, it lacks the comprehensive comparison over the conventional sequential models using the limited targets for contouring task. We not only address the inherent limitations of sequential models, but also showcase superior performance through extensive validation on prostate and head and neck cancer datasets. The introduction of the multi-task learning-based integrated radiotherapy framework suggests an important step toward more accurate, efficient, and integrated radiotherapy approaches to radiotherapy treatment planning.

\section{Methods}
\begin{figure}[!ht]
    \centering
    \includegraphics[width=\textwidth]{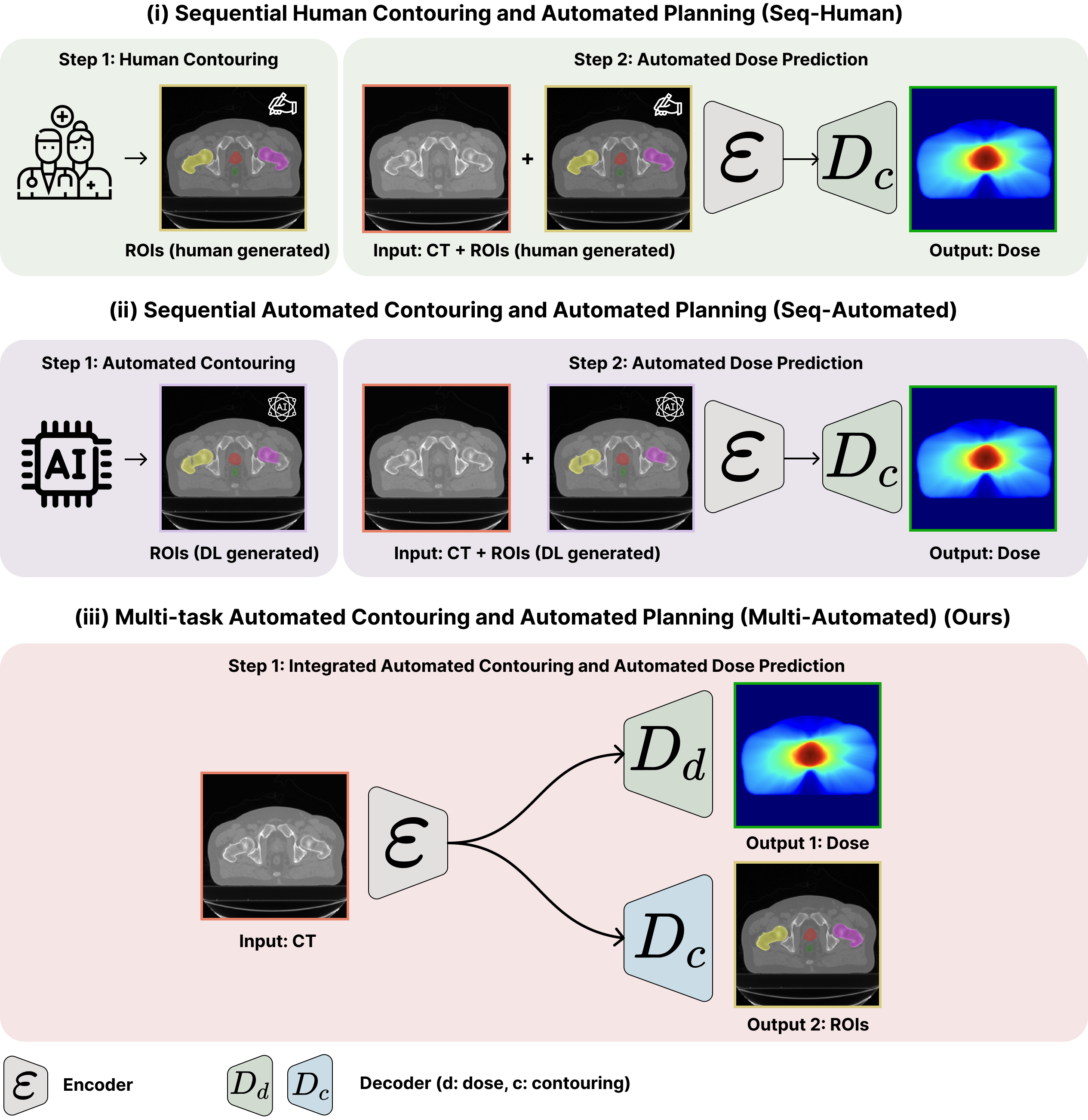}
    \caption{Architectures, inputs, and outputs of the (i) sequential human contouring and planning (Seq-Human), (ii) sequential automated contouring and planning (Seq-Automated), and proposed framework (iii) automated contouring and planning using multi-task learning (Multi-Automated). In Seq-Human and Seq-Automated, the input includes clinician labeled ground-truth (GT) and deep learning (DL) regions of interest (ROI), respectivley, both concatenated channel-wise with the CT image input. Multi-Automated uses only CT imaging as an input but includes an additional decoder for predicting ROI (output) simultaneously with dose distributions.}
    \label{fig:figure1}
\end{figure}

\subsection{Datasets}
To comprehensively evaluate our approach, we used datasets for two cancer treatment sites: prostate and head and neck. For both treatment sites, the datasets included the CT imaging, volumetric dose distributions generated from the radiotherapy plans, and the contoured anatomy of relevant regions of interest (ROIs).

The prostate cancer dataset was collected from 110 patients that underwent volumetric arc therapy (VMAT) at Princess Margaret Cancer Centre (Toronto, Canada) with a prescription dose of 60 Gy in 20 fractions. For prostate, contouring included five key ROIs: the prostate, rectum, bladder, left and right femur. The dataset was divided into training, validation, and test cohorts comprising 95, 5, and 10 patients, respectively. 

The head and neck dataset was collected from the publicly available OpenKBP Challenge dataset. We collected data from 328 patients who underwent radiotherapy, including radiotherapy treatment plans using either step-and-shoot intensity-modulated radiation therapy (IMRT) with nine approximately equispaced coplanar 6 MV fields. For head and neck, contouring included four ROIs: the brain stem, spinal cord, left and right parotid gland. as these four ROIs were consistently available across all data in the OpenKBP dataset. Following \cite{openkbp}, we split OpenKBP datasets into training, validation, and test sets of 200, 40, and 78 patients, respectively. We omitted 22 cases from the original OpenKBP dataset due to incomplete contour information for one or more target OARs.

We pre-processed all datasets to ensure the effective training DL models. For all datasets, we employed intensity clipping of CT to -1024 to 1500, followed by normalizing CT inputs by subtracting the mean and dividing by the standard deviation calculated from all CT scans in the training dataset. The mean and standard deviation of Hounsfield Unit in training datasets were -662.10 and 467.75 for Prostate, and 85.18 and 282.26 for OpenKBP dataset, respectively. We min-max normalized the target dose distribution maps by dividing with the prescribed dose for each dataset, which are 60 Gy and 70 Gy for the Prostate and OpenKBP dataset, respectively.

\subsection{Model architectures}
\subsubsection{Baseline.}
We developed a U-Net based model for dose prediction as the baseline model to compare with multi-task learning dose prediction. We designed an encoder-decoder architecture following the baseline model in the study by Kim \etal, which utilizes a ResNet-50 pre-trained backbone for the encoder \cite{kim2023cross}. The input for the baseline model is a 2-dimensional (2D) CT slice, and the output of this model is a paired 2D dose distribution map. We utilized the same architecture across different models. 

\subsubsection{Sequential dose prediction.} Sequential dose prediction models are DL models requiring both CT and contour information for dose prediction. We define two types of common sequential models: tasks manually (human) sequentially performed as in the conventional clinical process (Seq-Human), and tasks performed sequentially using DL an automated process (Seq-Automated). Both sequential models are trained to predict dose distributions with the CT and contours as inputs, with the distinction that Seq-Human uses manually defined human contours while Seq-Automated uses automated DL contours (details in \Fref{fig:figure1}). Thus, we trained sequential-conventional and sequential-automated models using contour information from distinct sources to see the impact of the quality of contours on DL dose prediction. Human contours for the Seq-Human model were obtained from the original datasets paired with CT and clinical dose distributions. To generate DL contours for the Seq-Automated model, we trained separate DL contouring models for each dataset using paired CT scans and human contours from the same datasets. The contouring models have the same architecture as the baseline dose prediction task except the output number of channels at the end of the model, which is replaced with the number of contouring labels as per dataset.

\subsubsection{Integrated multi-task learning.}
Multi-task learning algorithms enhance deep learning models by simultaneously training for multiple tasks, leveraging shared parameters to foster beneficial cooperation. This approach implicitly extracts additional training signals from related tasks within the existing dataset, rather than explicitly expanding the dataset. The shared components of the deep learning network are thought to be regularized by the various tasks, potentially improving model performance and generalization. However, current multi-task learning architectures in medical imaging often fall short in effectively sharing information across tasks, limiting potential performance gains. 
We propose an \textbf{integrated automated contouring and planning model using multi-task learning (Multi-Automated)} for simultaneous automatic contouring and dose prediction. Unlike sequential approaches, the integrated multi-task learning framework generates dose distribution maps and contours at the same time (\Fref{fig:figure1}). The primary advantage of this integrated framework is that a separate contouring process is not required. Specifically, we implemented the Multi-Automated model using cross-task attention network introduced in \cite{kim2023cross}, which maximizes the cross-task interaction for effectively training two tasks simultaneously. The Multi-Automated model comprises a single encoder for extracting common feature representations and two task-specific decoders for predicting dose distribution maps and contours. Cross-task attention network architectures will jointly train all models simultaneously with shared encoder components (i.e. feature representations) and per-task decoders with identical architecture across all tasks. Each task has its own attention mechanisms within the shared encoder, balancing per-task specialization with joint learning of cross-task convolution features. This design allows all models to extract the same base features while leveraging spatial relationships and feature interactions differently. During the training process, 2D CT images are fed into the shared encoders with cross-task specialized attention mechanisms, which then connect to the task-specific decoders. Further details on the cross-task attention network implementation can be found in \cite{kim2023cross}.

\subsubsection{Training Details.}
In training the dose prediction task, we adopted the mean absolute error (MAE) as our loss function, known for its effectiveness in regression models using medical imaging \cite{openkbp}. For training the contouring task of Multi-Automated and the baseline contouring model, we employed a combination loss function, known as combo loss, which integrates dice loss and cross-entropy loss through a weighted summation\cite{ma2021loss}. Herein, the weights for dice loss and cross-entropy loss were set to 0.3 and 0.7, respectively. This hybrid approach leverages the strengths of both loss functions to enhance model performance. 

To adaptively balance the weights of each loss function in the multi-task learning context, we employed the dynamic weight average technique, which significantly contributes to the stability and efficiency of model training\cite{liu2019end}. For training all models, we utilized batch size of 32 and 8 for the Prostate and OpenKBP datasets, respectively, ensuring optimal learning dynamics for each dataset's characteristics. Furthermore, we utilized the Adam optimizer with a learning rate of $10^{-4}$ and the weight decay of $10^{-5}$ for updating model parameters.

\subsection{Evaluation metrics}
\subsubsection{Dose prediction.}
In the evaluation of dose prediction accuracy for various ROIs in radiotherapy, we employed a comprehensive set of dose-volume metrics using clinical contours tailored to each anatomical region to ensure a robust and clinically relevant assessment and to maximize therapeutic outcomes while minimizing adverse effects. Following the clinical standard of care for prostate cancer radiotherapy treatment planning at Princess Margaret Cancer Centre, we evaluated the automated planning models using the dose volume histogram (DVH) metrics presented in \Tref{tab:metrics}.

\begin{table}[!ht]
\caption{Summary of dose-volume histogram (DVH) metrics used to evaluate dose prediction performance. DVH metrics assess critical aspects of radiotherpay treatment planning, including dose coverage, exposure, and volume thresholds for regions of interest (ROIs). Each metric is paired with the organs it is used to evaluate, providing a comprehensive overview of the criteria for analysis.}
\label{tab:metrics}
\begin{tabular*}{\textwidth}{@{}l*{15}{@{\extracolsep{0pt plus 12pt}}l}}
\br
DVH&Description&ROIs&\\
\mr
\hspace{1.0ex}$D_{mean}$&Mean dose, representing overall exposure&All&\\
\hspace{1.0ex}$D_{99}$&Dose received by 99\% of the volume&Prostate&\\
\hspace{1.0ex}$D_{50}$&Dose received by 50\% of the volume&Rectum, Bladder&\\
\hspace{1.0ex}$D_{30}$&Dose received by 30\% of the volume&Rectum, Bladder&\\
\hspace{1.0ex}$D_{5}$&Dose received by 5\% of the volume&Femur&\\
\hspace{1.0ex}$D_{0.1cc}$&Maximum dose received by&Brain stem, Spinal cord&\\
&the smallest volume of 0.1 cc&Left and Right parotid gland&\\
\hspace{1.0ex}$V_{30}$&Volume receiving at least 30 Gy&Rectum, Bladder&\\
\hspace{1.0ex}$V_{22}$&Volume receiving at least 22 Gy&Left and Right femur&\\
\hspace{1.0ex}$V_{14}$&Volume receiving at least 14 Gy&Left and Right femur&\\
\br
\end{tabular*}
\end{table}

\subsubsection{ROI Contouring.}
In this study, we quantitatively evaluated the contouring performance of ROIs using the Dice Score Coefficient and Hausdorff Distance, evaluation metrics widely regarded for its effectiveness in measuring the accuracy of image contouring\cite{ma2021loss}. The Dice coefficient compares the similarity between the predicted and the ground truth contours, providing a score between 0 and 1, where 1 indicates perfect agreement and 0 denotes no overlap. Mathematically, it is expressed as twice the shared information (intersection) between the predicted and actual contours, divided by the sum of pixels in both the predicted and ground truth contours. This ratio thus reflects both the size and location of segmented regions, making it an ideal measure for assessing contouring precision. On the other hand, the Hausdorff Distance serves as a complementary metric by calculating the maximum distance of the closest point from one contour to the other, effectively measuring the largest discrepancy between the predicted and ground-truth boundaries. This combination offers a robust evaluation of contouring performance, capturing both the overall accuracy and the extremities of prediction errors.

\subsubsection{Statistical test.}
For dose prediction, we tested the statistically significant differences of the MAE of DVH metrics (DVH-MAE) for our proposed method, Multi-Automated, with that of Seq-Automated model as per each DVH metrics across ROIs. We compared Seq-Automated with Multi-Automated, which are both fully-automated radiotherapy treatment planning models without any human intervention, but with different training strategy which are performed sequentially and integrated, respectively. We utilized one-sided paired t-tests ($\alpha$ = 0.05) for each DVH metric, corrected by Bonferroni correction for multiple comparisons. For contouring, we calculated statistical significant differences of Dice Score Coefficient and Hausdorff distance between Multi-Automated model compared to the baseline single-task learning contouring model for each ROI using one-sided paired t-tests, corrected by Bonferroni correction for multiple comparisons.
\section{Results}

\begin{table}[!ht]
\caption{Dose Volume Histogram - Mean Absolute error (DVH-MAE) for metrics (unit:Gy) for dose prediction models (i.e., Baseline model using CT as inputs (Baseline), Sequential human contouring and planning (Seq-Human), Sequential automated contouring and planning (Seq-Automated), and automated contouring and planning using multi-task learning (Multi-Automated)) for Prostate dataset. Relative differences of the DVH-MAE for each model are calculated compared to the baseline performance, positive value means DVH-MAE improvement compared to the baseline, and vice versa. Best results for each metric are bolded. $\dagger$ presents DVH-MAE scores for Multi-Automated with significant differences (p $<$ 0.05) compared to Seq-Automated, identified using paired t-tests and Bonferroni correction.}
\label{tab:dose_prostate}
\begin{tabular*}{\textwidth}{@{}l*{15}{@{\extracolsep{0pt plus
12pt}}l}}
\br
ROIs&Baseline&Seq-Human&Seq-Automated&Multi-Automated\\
/ DVH-MAE (Gy)&&&&(Ours)\\
\mr
Prostate &&&&\\
\hspace{2.0ex}$D_{99}$&2.530&\textbf{1.193}&1.738&2.284\\
\hspace{2.0ex}$D_{mean}$ &0.604&\textbf{0.427}&0.570&$1.177^{\dagger}$\\
Rectum &&&&\\
\hspace{2.0ex}$D_{30}$&7.236&\textbf{6.853}&7.822&7.493\\
\hspace{2.0ex}$D_{50}$&\textbf{5.538}&5.704&6.684&5.821\\
\hspace{2.0ex}$V_{30}$&\textbf{0.470}&0.489&0.551&0.476\\
\hspace{2.0ex}$D_{mean}$ &3.779&\textbf{3.716}&4.254&4.108\\
Bladder &&&&\\
\hspace{2.0ex}$D_{30}$&8.897&7.002&9.588&$\textbf{6.514}^{\dagger}$\\
\hspace{2.0ex}$D_{50}$&10.029&8.331&10.535&$\textbf{8.161}^{\dagger}$\\
\hspace{2.0ex}$V_{30}$&3.330&\textbf{2.609}&3.370&2.796\\
\hspace{2.0ex}$D_{mean}$ &7.153&\textbf{5.517}&7.360&5.876\\
Left femur &&&&\\
\hspace{2.0ex}$D_{5}$&3.738&4.127&4.151&\textbf{3.593}\\
\hspace{2.0ex}$V_{14}$&1.282&1.188&1.429&\textbf{1.059}\\
\hspace{2.0ex}$V_{22}$&2.386&3.121&3.311&\textbf{2.385}\\
\hspace{2.0ex}$D_{mean}$&3.044&3.305&3.460&\textbf{2.838}\\
Right femur &&&&\\
\hspace{2.0ex}$D_{5}$&3.754&4.825&4.739&$\textbf{3.683}^{\dagger}$\\
\hspace{2.0ex}$V_{14}$&1.082&\textbf{1.006}&1.453&1.030\\
\hspace{2.0ex}$V_{22}$&2.807&3.306&3.126&$\textbf{1.966}^{\dagger}$\\
\hspace{2.0ex}$D_{mean}$&2.523&2.817&3.278&$\textbf{2.254}^{\dagger}$\\
\mr
Average &3.900&3.641&4.301&\textbf{3.528}\\
Relative difference (\%)&&6.641&-10.28&\textbf{9.538}\\
\br
\end{tabular*}
\end{table}

\begin{table}[!ht]
    \caption{Dose Volume Histogram - Mean Absolute error (DVH-MAE) for metrics (unit:Gy) for dose prediction models (i.e., Baseline model using CT as inputs (Baseline), Sequential human contouring and planning (Seq-Human), Sequential automated contouring and planning (Seq-Automated), and automated contouring and planning using multi-task learning (Multi-Automated)) for OpenKBP dataset. Relative differences of the DVH-MAE for each model are calculated compared to the baseline performance, positive value means DVH-MAE improvement compared to the baseline, and vice versa. Best results for each metric are bolded. $\dagger$ denotes DVH-MAE scores for Multi-Automated, highlighting significant differences (p $<$ 0.5) compared to Seq-Automated, determined through paired t-tests and Bonferroni correction.}
    \label{tab:dose_openkbp}
    \begin{tabular*}{\textwidth}{@{}l*{15}{@{\extracolsep{0pt plus
    12pt}}l}}
    \br
    ROIs &Baseline&Seq-Human&Seq-Automated&Multi-Automated\\
    / DVH-MAE (Gy)&&&&(Ours)\\
    \mr
    Brain stem &&&&\\
    \hspace{2.0ex}$D_{0.1cc}$&\textbf{18.340}&28.388&29.610&29.922\\
    \hspace{2.0ex}$D_{mean}$ &\textbf{3.098}&4.227&3.467&3.634\\
    Spinal cord &&&&\\
    \hspace{2.0ex}$D_{0.1cc}$&25.766&29.890&31.571&\textbf{21.748}\\
    \hspace{2.0ex}$D_{mean}$ &\textbf{2.191}&2.481&2.523&2.438\\
    Left parotid &&&&\\
    \hspace{2.0ex}$D_{0.1cc}$&5.092&5.184&5.379&\textbf{4.573}\\
    \hspace{2.0ex}$D_{mean}$&8.068&7.674&8.062&\textbf{7.193}\\
    Right parotid &&&&\\
    \hspace{2.0ex}$D_{0.1cc}$&5.236&5.211&4.563&\textbf{4.279}\\
    \hspace{2.0ex}$D_{mean}$ &7.699&7.722&8.025&$\textbf{7.081}^{\dagger}$\\
    \mr
    Average &9.436&11.347&11.650&\textbf{10.109}\\
    Relative difference (\%)&&-20.252&-23.463&-\textbf{7.132}\\
    \br
    \end{tabular*}
\end{table}

\subsection{Dose prediction}
The DVH-MAE in dose prediction of each model type for the Prostate and OpenKBP datasets is reported in \Tref{tab:dose_prostate} and \Tref{tab:dose_openkbp}, respectively. Our proposed Multi-Automated model improved the average performance of dose prediction compared to the baseline and the two sequential models (Seq-Human, using human contoured ROIs and CT as inputs, and Seq-Automated, using AI contoured ROIs and CT as inputs). These trends hold consistently across both datasets, indicating that Multi-Automated improves dose prediction performance irrespective of cancer type or anatomical site. Specifically, in \Tref{tab:dose_prostate}, for the prostate dataset, Multi-Automated achieves a DVH-MAE of 3.528 Gy, outperforming the baseline single-task learning model and the two sequential models, which scored 3.900 Gy, 3.641 Gy, and 4.301 Gy, respectively. Similarly in \Tref{tab:dose_openkbp}, for the head and neck cancer dataset, Multi-Automated achieves an MAE of 10.109 Gy, while the baseline single-task learning model and sequential models score 9.436 Gy, 11.347 Gy, and 11.650 Gy, respectively. 

\begin{figure}[!ht]
    \centering
    \includegraphics[width=\textwidth]{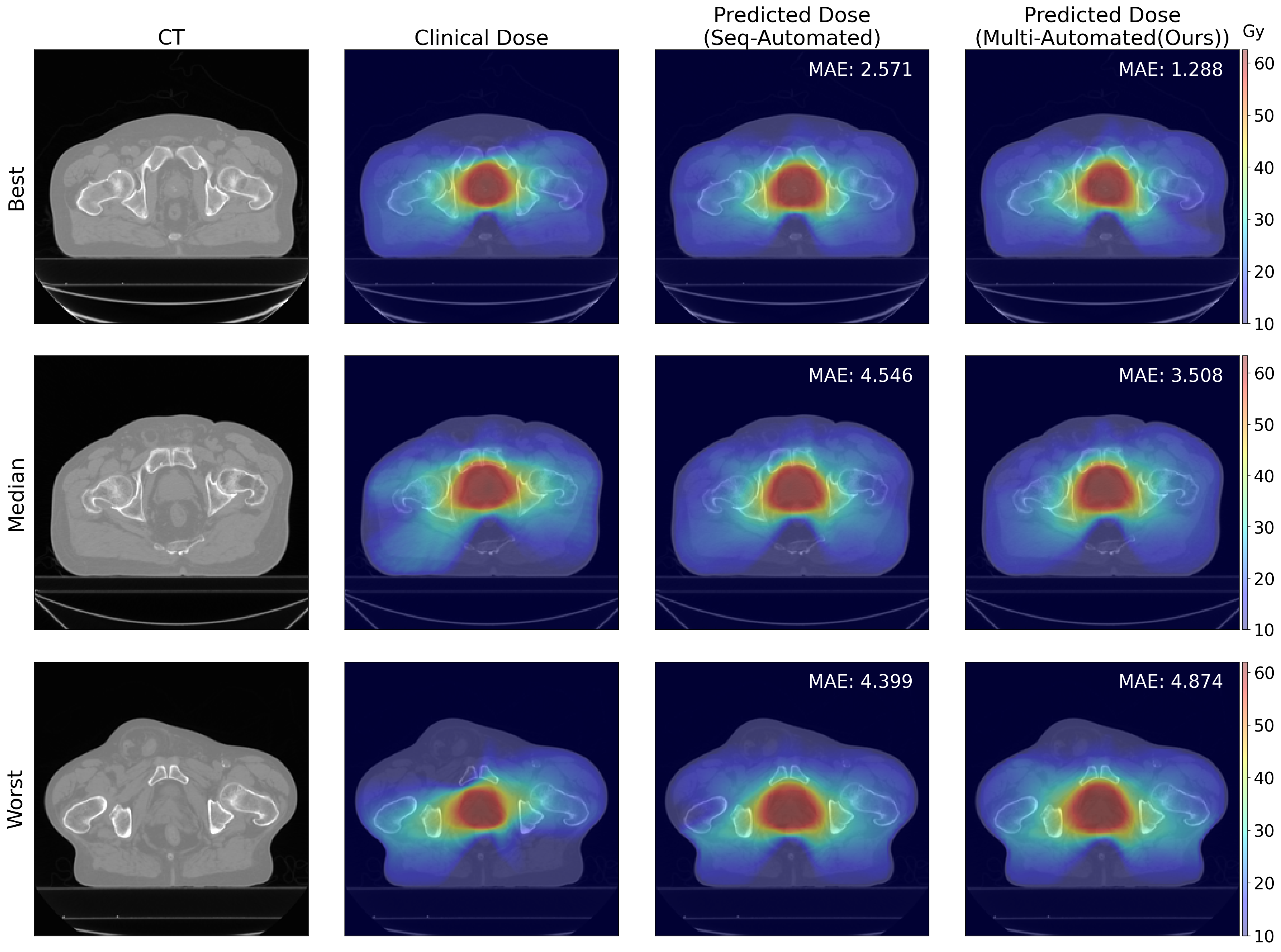}
    \caption{This figure illustrates qualitative results of Prostate dataset, showing input CT scans, ground truth clinical dose distributions (Clinical Dose), and the dose distributions generated by sequential automated contouring and planning model (Seq-Automated) and the automated contouring and planning model using multi-task learning (Multi-Automated). The voxel-wise Mean Absolute Error (MAE) for both model outputs are provided. The figure presents best, median, and worst cases in terms of MAE from the test dataset}
    \label{fig:dose_prostate}
\end{figure}

Furthermore, we computed relative difference in MAE to compare the performance of the baseline model with the two sequential models, Seq-Human and Seq-Automated. For the prostate dataset, Seq-Human outperformed the baseline model by 6.641\% while Seq-Automated performed 10.28\% worse than the baseline. This result demonstrates the quality of the DL ROIs are worsening the dose prediction performance when they are imperfectly contoured. For the OpenKBP dataset, however, baseline models outperformed both sequential models, Seq-Human and Seq-Automated, by 20.252\% and 23.463\%, respectively, note that the OpenKBP dataset is not originally designed for training contouring models, meaning that the provided contour information might not be perfectly curated. Meanwhile the relative difference of Multi-Automated is 7.132\% worse than the baseline, which is still outperforming both sequential models with the large margin. However, Multi-Automated outperformed $D_{0.1cc}$ and $D_{mean}$ for right and left parotid, meaning that Multi-Automated better predicts the dose distribution for OARs. In \Fref{fig:dose_prostate} and \Fref{fig:dose_openkbp}, we visualized the dose prediction results using the prostate dataset, and OpenKBP dataset, respectively.

To validate the impact of multi-task learning for developing dose prediction models in automated radiotherapy, we compared the DVH-MAE scores of Seq-Automated with Multi-Automated which are both used without any human interventions. The average DVH-MAE of Seq-Automated and Multi-Automated were 4.301 Gy and 3.528 Gy for Prostate dataset, respectively. Especially the improvement of using Multi-Automated were significant (p $<$ 0.05) in critical anatomical structures; Bladder ($D_{30}$ and $D_{50}$) and in Right femur ($D_{5}$, $V_{22}$, and $D_{mean}$). For OpenKBP dataset, DVH-MAE of Seq-Automated and Multi-Automated were 11.650 Gy and 10.109 Gy, respectively, where $D_{mean}$ of right parotid glands were significantly improved for Multi-Automated (p $<$ 0.05).

Although the D99 metric reveals sub-optimal performance in prostate gland contouring, this may be attributed to the inherent trade-offs between the contouring and dose prediction. Nevertheless, the capability of Multi-Automated model to predict dose distributions while being aware of ROIs during training allows it to better optimize dose distributions by minimizing exposure to OARs. This is evident in its improved DVH-MAE scores for various organs, including the bladder, left and right femur, prostate, spinal cord, and parotid glands, as shown in \Tref{tab:dose_prostate} and \Tref{tab:dose_openkbp}. By simultaneously training the Multi-Automated model for contouring ROIs and dose prediction tasks, the increased awareness of the model can ultimately improve radiotherapy treatment planning.

\subsection{Contouring}
\begin{figure}[!ht]
    \centering
    \includegraphics[width=\textwidth]{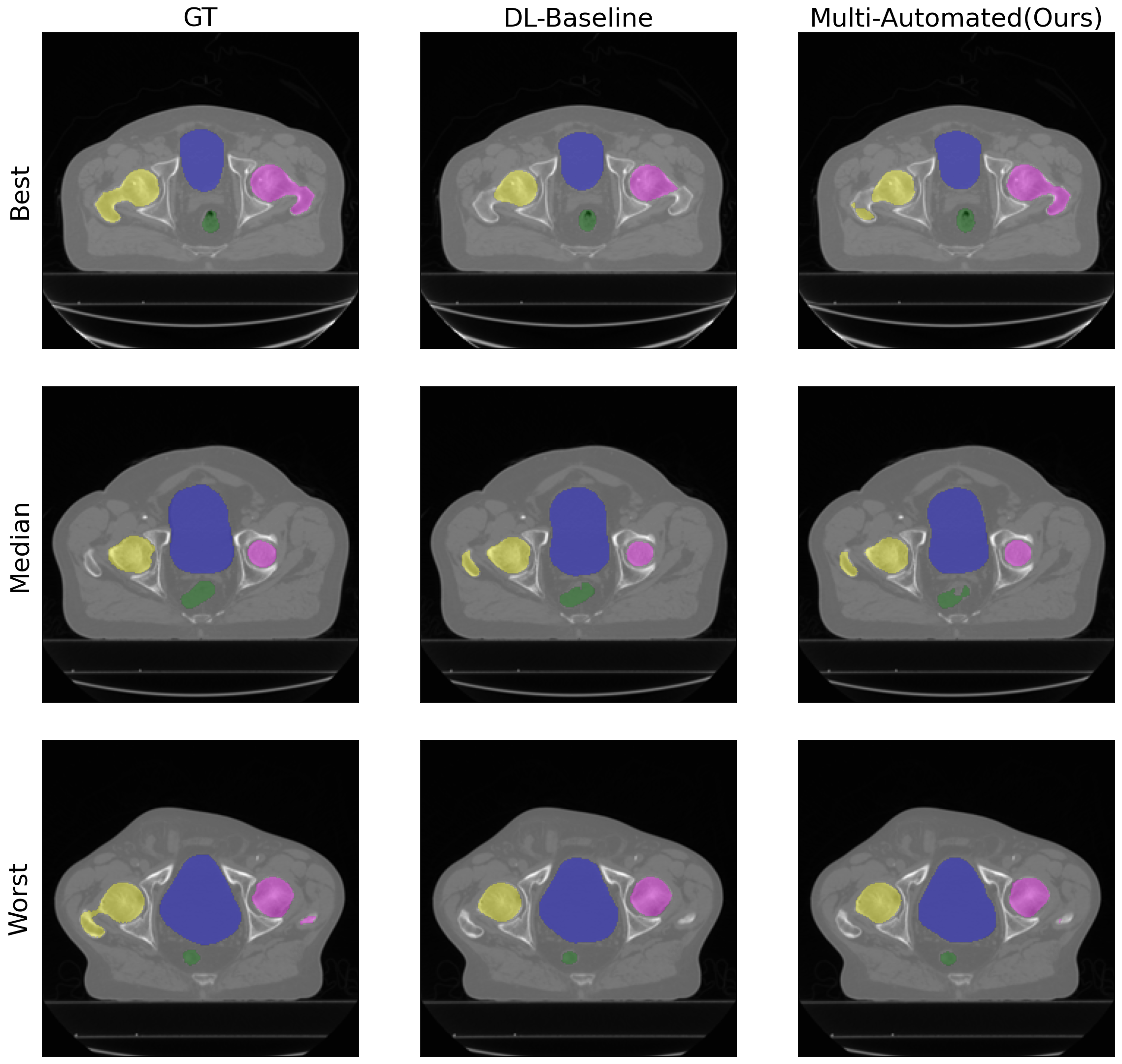}
    \caption{In this figure, we present the results of automated DL contouring using Prostate dataset. The image showcases the ground truth labels (GT), and the DL contouring outputs from the baseline model (DL-Baseline) and the automated multi-task contouring and planning model (Multi-Automated). The figure includes representative examples, displaying best, median, and worst cases from our test dataset.}
    \label{fig:seg_prostate}
\end{figure}

\begin{figure}[!ht]
    \centering
    \includegraphics[width=\textwidth]{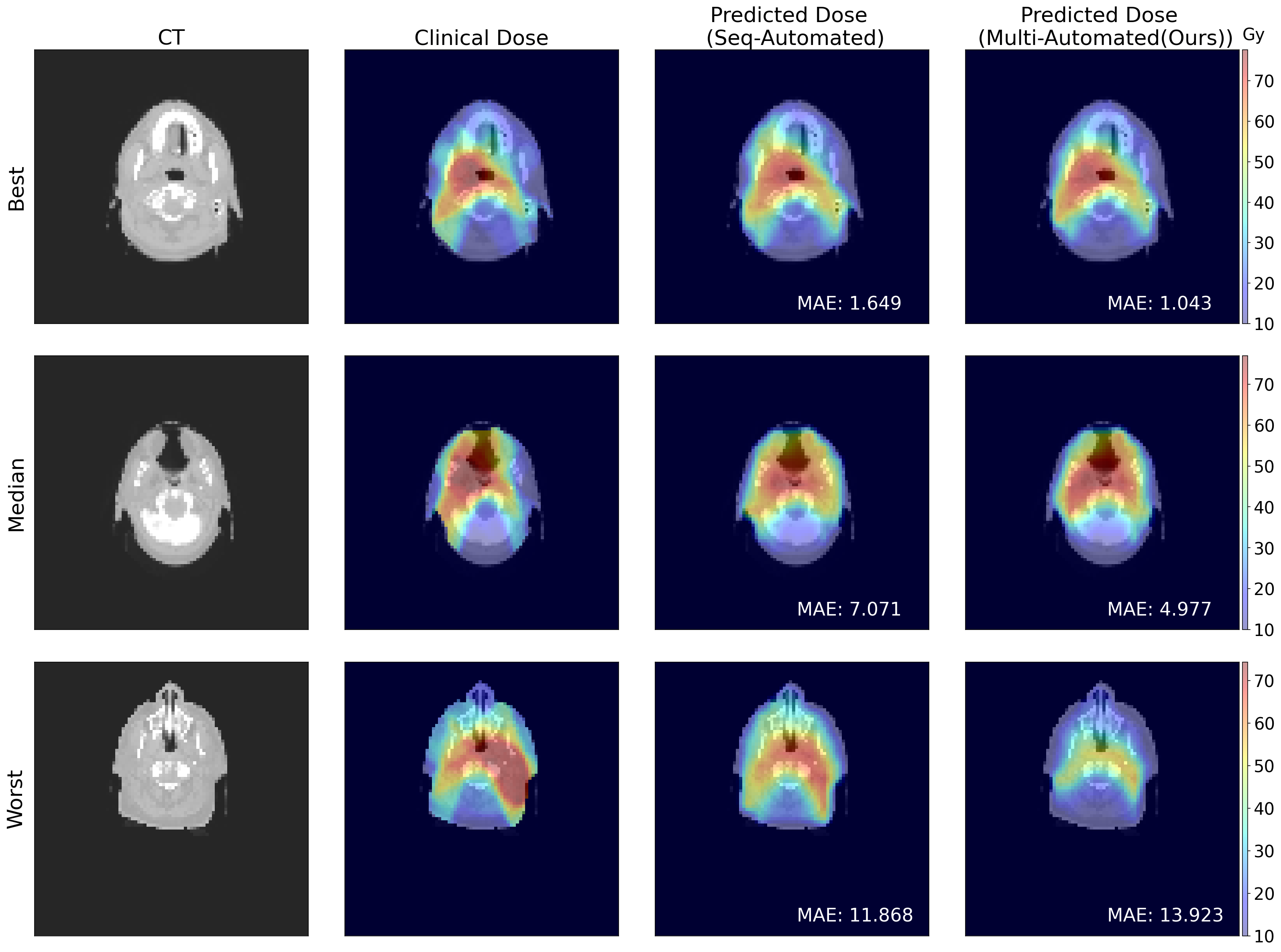}
    \caption{This figure presents qualitative results from the OpenKBP dataset, focusing on the head and neck cancer, showing input CT scans, ground truth clinical dose distributions (Clinical Dose), and the dose distributions generated by sequential automated contouring and planning (Seq-Automated) and the automated contouring and planning model using multi-task learning (Multi-Automated). The Mean Absolute Error (MAE) for both model outputs are provided. The figure presents the best, median, and worst cases from the test dataset, starting from the top.}
    \label{fig:dose_openkbp}
\end{figure}

\begin{figure}[!ht]
    \centering
    \includegraphics[width=\textwidth]{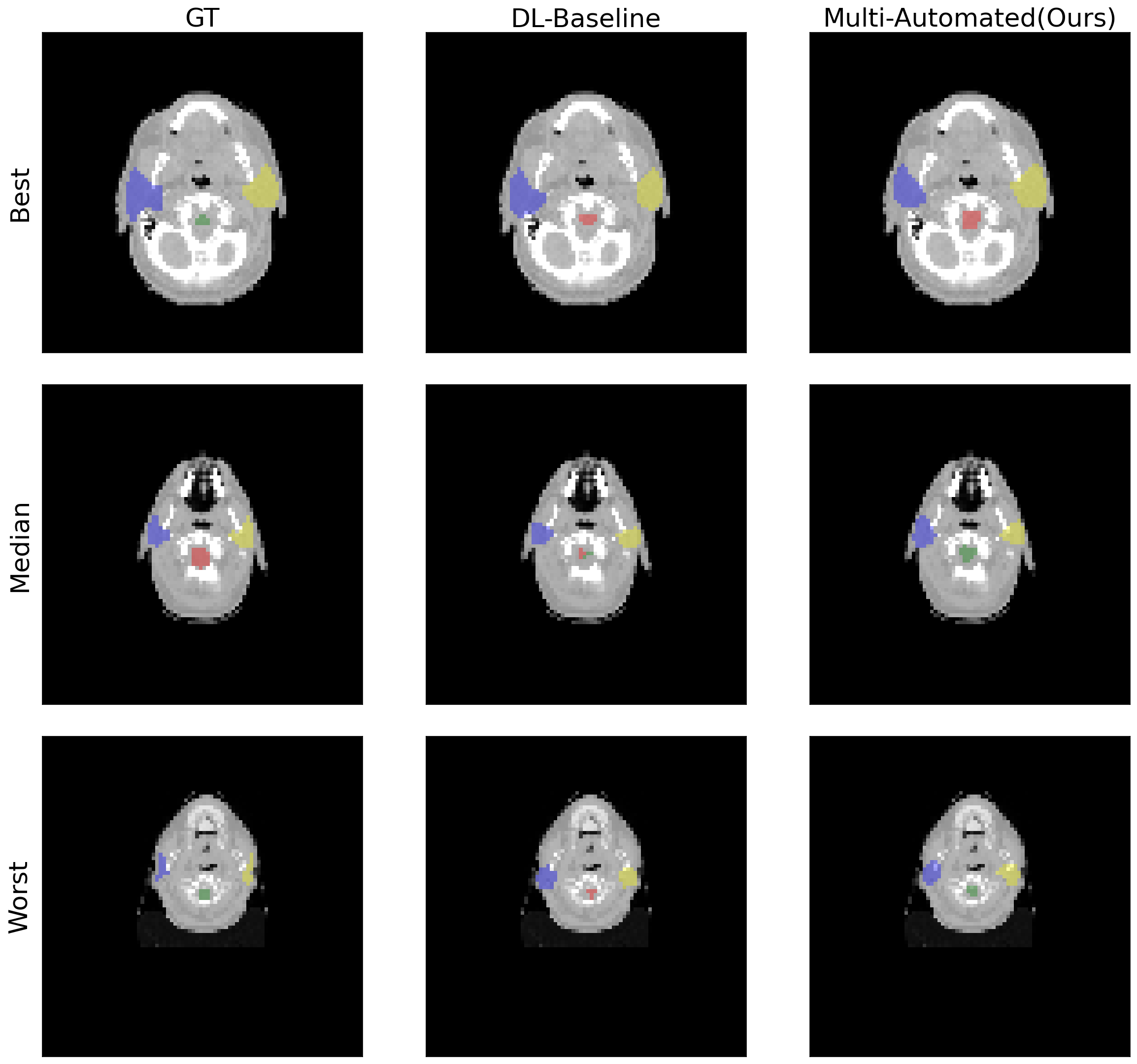}
    \caption{In this figure, we present the results of DL automated contouring using OpenKBP dataset. The image showcases the ground truth labels (GT), and the automated contouring outputs from the baseline (DL-Baseline) and the automated contouring and planning model using multi-task learning (Multi-Automated). Additionally, we provide the Dice Score Coefficient for the segmentation output. The figure includes representative examples, displaying the best, median, and worst cases from our test dataset.}
    \label{fig:seg_openkbp}
    \label{fig:figure5}
\end{figure}

Additionally, results in \Tref{tab:seg_prostate} and \Tref{tab:seg_openkbp} show that Multi-Automated achieves comparable performance in contouring when trained for dose prediction simultaneously. The Dice Score Coefficients for the baseline single-task learning model for contouring are 0.818 and 0.674 for the prostate and OpenKBP datasets, respectively. In contrast, Multi-Automated achieves dice score coefficients of 0.824 and 0.716, respectively, showing better performance in OpenKBP datasets. For Hausdorff distance, compared to 7.549 and 39.831 in baseline, Multi-Automated results show 12.049 and 22.872 for Prostate and OpenKBP datasets, respectively. In\ \Fref{fig:seg_prostate} and \Fref{fig:seg_openkbp}, we further illustrate the qualitative results of both the baseline model and Multi-Automated for the prostate and OpenKBP datasets, respectively. 

\begin{table}[!ht]
    \caption{Dice score coefficient and Hausdorff distance using Prostate dataset for baseline DL contouring model (DL-Baseline) and the automated contouring and planning model using multi-task learning (Multi-Automated). Better results for each ROI and each metric are bolded. Higher the better for dice score coeffcient, denoted as $\uparrow$, lower the better for Hausdorff distance, denoted as $\downarrow$. $\dagger$ denotes both contouring metrics for Multi-Automated, highlighting significant differences (p $<$ 0.5) compared to baseline, determined through paired t-tests and Bonferroni correction.}
    \label{tab:seg_prostate}
    \begin{tabular*}{\textwidth}{@{}l*{15}{@{\extracolsep{0pt plus 12pt}}c}}
    \br
    ROIs&\multicolumn{2}{c}{Dice score coefficient($\uparrow$)}&\multicolumn{2}{c}{Hausdorff Distance ($\downarrow$)}\\
    &\multicolumn{1}{c}{DL-Baseline}&\multicolumn{1}{c}{Multi-Automated}&\multicolumn{1}{c}{DL-Baseline}&\multicolumn{1}{c}{Multi-Automated}\\
    \mr
    Prostate &\textbf{0.839}&0.836&6.521&\textbf{2.900}\\
    Rectum &0.807&\textbf{0.816}&\textbf{5.469}&5.603\\
    Bladder &\textbf{0.796}&0.795&\textbf{4.593}&4.849\\
    Left femur &\textbf{0.822}&0.819&11.491&\textbf{10.452}\\
    Right femur &0.827&\textbf{0.854}&\textbf{9.669}&$36.441^{\dagger}$\\
    \mr
    Average &0.818&\textbf{0.824}&\textbf{7.549}&12.049\\
    \br
    \end{tabular*}
\end{table}

\begin{table}
    \caption{Dice score coefficient and Hausdorff distance using OpenKBP dataset for both DL contouring models: baseline (DL-Baseline) and the automated contouring and planning model using multi-task learning (Multi-Automated). Better results for each ROI and each metric are bolded. Higher the better for dice score coeffcient, denoted as $\uparrow$, lower the better for Hausdorff distance, denoted as $\downarrow$. $\dagger$ Denotes  significant differences (p $<$ 0.5) in either contouring metric compared to baseline, determined through paired t-tests and Bonferroni correction.}
    \label{tab:seg_openkbp}
    \begin{tabular*}{\textwidth}{@{}l*{15}{@{\extracolsep{0pt plus 12pt}}c}}
    \br
    ROIs&\multicolumn{2}{c}{Dice score coefficient($\uparrow$)}&\multicolumn{2}{c}{Hausdorff Distance ($\downarrow$)}\\
    &\multicolumn{1}{c}{DL-Baseline}&\multicolumn{1}{c}{Multi-Automated}&\multicolumn{1}{c}{DL-Baseline}&\multicolumn{1}{c}{Multi-Automated}\\
    \mr
    Brain stem &0.717&$\textbf{0.750}^{\dagger}$&46.103&$\textbf{22.073}^{\dagger}$\\
    Spinal cord &0.628&$\textbf{0.716}^{\dagger}$&\textbf{13.981}&16.002\\
    Left parotid &0.673&$\textbf{0.693}^{\dagger}$&44.679&$\textbf{30.106}^{\dagger}$\\
    Right parotid &0.680&$\textbf{0.704}^{\dagger}$&54.563&$\textbf{23.308}^{\dagger}$\\
    \mr
    Average &0.674&\textbf{0.716}&39.831&\textbf{22.872}\\
    \br
    \end{tabular*}
\end{table}


\section{Discussion}
Our study introduces an integrated multi-task learning framework for automatic contouring and dose prediction for the purposes of radiotherapy treatment planning. 
The primary contribution of our study lies in demonstrating the efficacy of multi-task learning in the context of treatment planning. We showed that integrating contouring and dose prediction tasks can lead to improved accuracy and efficiency in generating treatment plans. This integration allows the model to more comprehensively understand the patient anatomy, ultimately leading to enhanced dose prediction accuracy. The successful application of our approach in two distinct treatment sites, prostate and head and neck, further underscores the versatility and potential of the proposed framework for wide-ranging clinical applications.

One of the key strengths of our approach is its robustness against the variability of input contourings, a common challenge in sequential DL models for automated radiotherapy treatment planning. By learning to contour while at the same time predicting dose distributions, our model reduces the dependency on the quality of input contours, which is a significant improvement over conventional sequential methods. This highlights that model training in the integrated framework utilizes the inherent correlations between contouring anatomical structures and predicting dose distributions, thereby improving the robustness and accuracy of the treatment planning process.

To exhibit the impact of contour quality, we compared Seq-Human and Seq-Automated in~\Tref{tab:dose_prostate} and \Tref{tab:dose_openkbp} and found a performance gap between the two sequential models that predict dose distributions based on human and automated contours with CT images as inputs. On the other hand, the results of our proposed Multi-Automated model ensure that DL-based automated radiotherapy remains effective regardless of the quality of contours by simultaneously learning to segment the ROIs, a feature that supports its practicality and reliability for clinical settings~\cite{gu2023doseinputdata}. In addition, Multi-Automated showed comparable contouring performance in both prostate and head and neck sites. Moreover, results in \Tref{tab:seg_prostate} and \Tref{tab:seg_openkbp} showcase that the Multi-Automated sometimes outperformed baseline contouring performance. This indicates that the multi-task learning framework can achieve better dose prediction performance without compromising contouring performance, sometimes even improving contouring performance.

However, our study is not without limitations. Although our proposed Multi-Automated model demonstrated the feasibility of simultaneous automated contouring and treatment planning, the results are limited to the certain ROIs for both sites. This need to be further validated under more clinically relevant situations. Moreover, the predicted dose distributions are not deliverable plans. However, the predicted dose distributions from Multi-Automated model can be used as the input to generate fully deliverable plans, a capability rarely addressed in most dose prediction studies~\cite{gronberg2023deep_seq1, nguyenpmb3d_seq4, gao2023flexible_seq5}.

Furthermore, one critical aspect of our approach is its current inability to control the dose outputs using input contours as effectively as the conventional sequential models can. While Multi-Automated improves robustness and performance for both tasks, the lack of direct control over dose distributions based on input contours may be a notable drawback. However, our findings also highlight the inherent trade-off between the improved generalizability and precision offered by multi-task learning approaches like Multi-Automated, which may sacrifice some degree of fine-tuned control over dose distribution maps based on input contours to achieve better overall performance in the two treatment planning tasks. This compromise underscores the need for careful consideration of the specific clinical contexts and treatment goals when selecting a method for radiotherapy treatment planning. We plan to further refine the Multi-Automated framework to accommodate variations in contour quality without compromising the model's ability to guide and adjust dose distributions effectively. This advancement would not only enhance the clinical viability of our approach but also extend its applicability to a broader range of treatment scenarios and complexities. Another aspect is that even though we developed the framework for integrated Multi-Automated model for efficient radiotherapy processes, its scalability is limited in terms of the need to train separate models for different treatment sites. As a future step toward generalization, we plan to further integrate different treatment sites into a single multi-task multi-modal model.

\section{Conclusion}
In conclusion, our study presented the development and validation of a novel multi-task learning framework specifically designed for two crucial tasks in radiotherapy treatment planning: contouring and dose prediction. We achieved substantial improvements in the overall performance of both tasks compared to models trained separately. This advancement is particularly promising in mitigating the variability associated with human-generated contouring, thereby contributing to the precision and reliability of DL automated radiotherapy systems. Our research highlights the potential of the integrated multi-task learning for automated contouring and treatment planning, paving the way for more accurate and efficient radiotherapy.
\newcommand{\newblock}{}
\bibliographystyle{unsrt}
\bibliography{reference}

\end{document}